**Role of Van der Waals forces in graphene adsorption over Pd, Pt and Ni.**

**Abstract**

We report *ab initio* computations with the Vienna ab Initio Simulation Package (VASP) aimed at elucidating the adsorption mechanism of graphene-like structures on (111) Pd, Pt, and Ni surfaces. To study the adsorption properties, we simulate an already-formed graphene layer. We present a comparative discussion of the graphene interactions with the three metals, focusing on the very particular adsorption of graphene over Pd.

**Introduction**

Since the discovery of graphene, in 2004 [1], "how to make graphene grow easily" has been an important research topic. Advances along that line of work have brought special attention to the nature of the interaction between graphene and metals, examples of substrate elements favoring chemical or physical bonding being found. A glance at this aspect of the periodic table shows that graphene adsorption on Pd lies at the frontier separating physisorption from chemisorption.

For certain metal-graphene systems the π-band lies substantially lower than in free graphene, so that a gap between the valence band and the conduction band opens around the K point of the Brillouin Zone, pushing the system to a semiconducting state; in other metal-graphene systems, however, the characteristic bands display no significant displacement [2, 3].

In recent years Giovannetti et al. [4], Khomyakov et al. [5] and Leibo Hu et al. [6] have studied the graphene adsorption over different metals. Giovannetti et al. [4] affirm that graphene adsorption over Pd is chemisorption, because the mean distance $Z$ between the graphene layer and substrate is less than 2.3A. That the Dirac-cone distortion of the electronic structure at the Fermi level argues in favor of chemisorption was pointed out by Khomyakov et al. [5]. Leibo Hu et al. [6], however, sustain that while graphene is chemisorbed to a single Pd atom, its bonding to a Pd surface is physical. In another study, Wintterlin and Bocquet [7] propose a new criterion, which labels the adsorption as physical if the graphene is separated by an equilibrium distance of 3.35 A.

As widely known, graphite is a set of graphene layers kept together by the weak Van der Waals forces. Here, we gauge the importance of those forces in graphene adsorption on Pd, Pt, and Ni substrates. Until recently, Van der Waals interactions could not be reliably treated with the VASP code; due to this limitation, the literature showed no consensus on the most suitable exchange-correlation functional to describe graphene-substrate interactions. Since 2010, however, when version 5.2.8 was released, the VASP code has included a Van der Waals semi-empirical correction based on Grimme's method [8]. We have therefore carried out a comparative study of three alternative functionals: the LDA, GGA, and GGA+VdW. Our results show that the present implementation of the Van der Waals forces is sufficient to settle the discussion.

2. Computational details

Our Density Functional calculations were carried with the *Vienna ab Initio Simulation Package* (VASP) [9-10]. The Kohn-Sham one-electron wave functions were expanded on a basis of plane waves with a cutoff value of 500 eV for the kinetic energy. The exchange-correlation functional was treated according to the Generalized Gradient Approximation (GGA) in the Perdew–Wang parametrization (PW91) [11]. A few of the calculations were performed in the Local Density Aproximation (LDA) [12]. Previous studies [13] and our tests have revealed no noticeable spin-polarization effects in either the Pd, Pt substrates or the atoms adsorbed on them. Therefore, except for the case of a free carbon atom on a Ni surface, all calculations were spin-restricted.

The interaction between atomic cores and valence electrons was described by the projector augmented wave (PAW) method [14,15]. The blocked Davidson approach was applied as the electronic minimization algorithm. We used the Monkhorst-Pack $k$-point mesh [16] and the Methfessel-Paxton technique [17], with an electronic-level smearing factor of 0.2. To choose the dimensions of the $k$-point mesh, we increased the mesh until the energy converged with better than 1 meV/atom accuracy. A $5\times5\times1$ k-point mesh resulted.

Structures were optimized until the maximum force on each atom became smaller than 10 meV/A. The electronic structure of the adsorbed C was analyzed on the basis of the Density of States (DOS).

To carry out the semi-empirical Van der Waals correction we applied the Grimme method [8], implemented in the VASP code since version 2.5.8. For Pd and Ni, the default parameters were used: $R_{VdW}$ = 30 A, d = 20 A, $C_6$ = 24.67, and $R_{Pd}$ = 1.639 A for Pd, and $R_{VdW}$ = 30 A, d = 20 A, $C_6$ = 10.80, and $R_{Pd}$ = 1.562 A for Ni. The parameters for Pt the parameters were fixed at the values $R_{VdW}$ = 12 A, d = 20 A, $C_6$ = 1.75, and $R_{Pt}$ = 1.452 A.

3. Slab models

An FCC stacking layered structure was assumed for the calculations. The reference M(111) (M = Pd, Pt and Ni) surface was represented by four-layer slabs. Preliminary tests with slab models containing up to five atomic layers have shown that four layers suffice to insure convergence, the surface-energy difference between four- and five-layer slabs being close to 0.005 eV. The repeated-atom slabs were separated in the $z$ direction by a vacuum region equivalent to five interlayer spaces optimized to avoid the interaction between them. Atom positions in the bottom three layers were kept frozen as optimized for the M bulk, whereas the other layer, closer to the adsorbate, was allowed to completely relax within a maximum-force criterion of 0.01 eV/A, which provided an interatomic distance relaxation of 1.5 % or less. The calculated cell parameters came to 2.79 A for Pd, 2.82 for Pt, and 2.50 A for Ni. As expected, the calculated GGA bond separations are somewhat larger than the corresponding experimental values of 2.75 A (Pd), 2.72 A (Pt), and 2.49 A (Ni).

*4.1 Study of graphene monolayer with different exchange and correlation functional.*

As a first step, we have modeled a graphene monolayer (G) adsorbed over the three metals Pd, Pt and Ni, as depicted in Fig. 1, and on another graphene monolayer. In all cases the mismatches were small, never greater than 2%. We have simulated an adsorbed graphene layer over Pd and Pt with a √3x√3 unit cell, which corresponds to θ = 8/3 coverage, and over Ni with a 2x2 unit cell, corresponding to θ = 2 coverage. The results of dynamic relaxation computations with different exchange-correlation functionals are presented in Table 1.

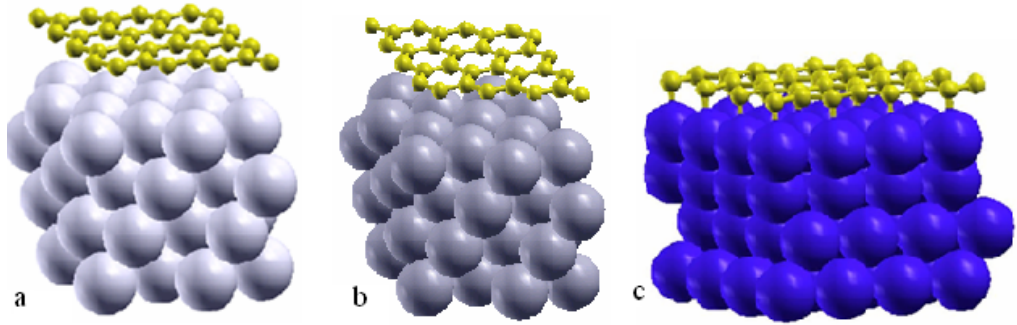

**Figure 1.** (Color online) Models for a graphene monolayer on (**a**) Pd, (**b**) Pt, and (**c**) Ni.

**Table 1**. Equilibrium distance Z (A) and $E_{ad}$ (kJmol$^{-1}$) for graphene over graphene (G-G), and graphene over Pd, Pt and Ni (G-Pd, G-Pt and G-Ni). The computations were carried out in the LDA, GGA and GGA+VdW.

| Aproximation | G-G | | G-Pd | | G-Pt | | G-Ni | |
|---|---|---|---|---|---|---|---|---|
| | Z | $E_{ads}$ | Z | $E_{ads}$ | Z | $E_{ads}$ | Z | $E_{ads}$ |
| LDA | 3.30 | -3 | 2.55 | -6 | 3.30 | -1 | 2.00 | -23 |
| GGA | 3.45 | 0* | 3.43 | 0* | 3.80 | -2 | 2.10 | -2 |
| GGA+VdW | 3.05 | -60 | 2.75 | -11 | 3.65 | -6 | 2.07 | -15 |
| Experimental | 3.35[a] | | - | | 3.70[b] | | 2.10[c] | |

* small, negative value, very close to zero. [a][7], [b][19], [c][20].

For comparison, we also modeled graphene on graphene, a typical physisorption system. The distance between graphene layers in graphite is 3.35 A [7]. The LDA result in Table 1 agrees well with this experimental datum. By contrast, the distances predicted by LDA for the graphene-metals systems are smaller than the experimental values of 3.70 A for G-Pt [18], and 2.10 A for G-Ni [19]. The opposite resulted from the GGA computations, which overestimates the measured distances for G-G and G-Pt. Finally, when Van der Waals forces were included in calculations, the resulting separations agreed well with experiment.

*4.2 Graphene adsorption over Pd, Pt and Ni.*

We have modeled an already-formed graphene layer adsorbed over different systems. To calculate the adsorption energy for given distance *Z*, we froze the sytem geometry. Three different exchange-functionals were tried: the GGA, LDA, and GGA +VdW. The results are shown in Fig. 2.

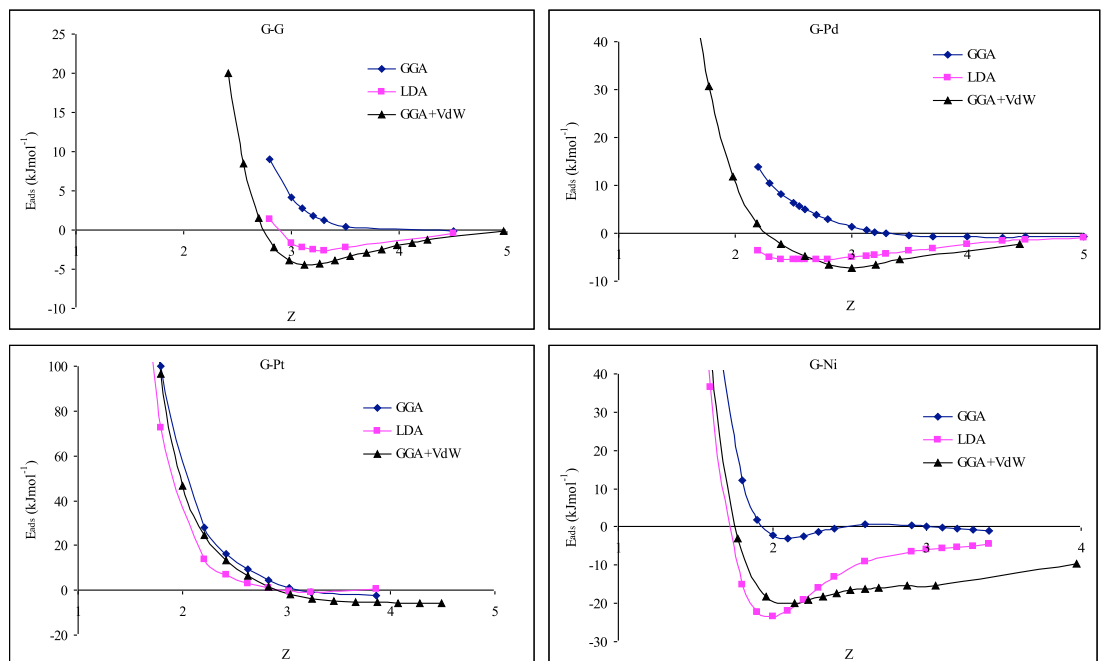

Figure 2. (Color online) Adsorption energies from the three indicated exchange-correlation functionals, GGA, LDA, and GGA+VdW, as a function of the separation *Z*. In clockwise order from top-left, the panels display the results for graphene (G) adsorption over graphene, Pd, Pt, and Ni.

The optimized adsorption energies and corresponding geometries are reported in Table 1. In Fig. 2, the calculation is not dynamic, neither the C atoms nor the metal surface atoms being allowed to relax. This explains the differences between the distances *Z* in Table 1 and the equilibrium *Z*'s in Fig. 2.

For the intensively-studied case of graphite on graphite (G-G, solid rhombi in the top-left panel of Fig. 2), in contrast with the LDA and LDA+VdW, the GGA functional curve has no discernible minimum; this indicates that the GGA functional is inadequate to describe the interaction between graphene layers.

For G-Pd (top-right panel of Fig. 2), once again the GGA functional shows no minimum comparable to the ones in the LDA and GGA-VdW curves. Similar to those in the top-left panel, the curves suggest that the adsorption over Pd is physical.

The G-Pt system (Fig. 2, bottom-left panel) shows no significant minimum for any of the three functionals. The low diffusion barrier points to no adsorption of the graphene monolayer over Pt. Experimental studies of graphene structures over Pt report irregular growth [20], which may be due to the weak interactions.

The G-Ni system (Fig. 2, bottom-right panel) is a clear example of graphene chemisorption, as indicated by the three curves. Neither the energies nor the geometries resulting from the LDA agree with previous reports [21].

The significant differences among the results obtained with different functionals highlights the importance of the Van der Waals corrections to the GGA. For example, the GGA results for G-G denies the existence of graphite. The Grimme method (GGA+VdW) had already been shown to improve the agreement with experiments [22,23]. The improvement over previous reports found in our study attests to the importance of the VdW upgrade in the VASP code.

**Conclusions**

Graphene monolayer-adsorption on Pd shows particularities separating it from adsorptions on other transitions metals. Our results shows that the Van der Waals forces must be taken into account, just as they must be included in analyses of the interaction between graphite layers.

For graphene on Pd, Fig. 2 shows that the LDA and GGA+VdW functionals yield markedly incongruent energy vs. distance plots, even though both exhibit clear minima. For G-Pt and G-Ni, by contrast, the energies calculated in the LDA are close to those computed with the GGA+VdW functional.

It is hard to decide whether graphene is physisorbed or chemisorbed over Pd, because the mean distance between graphene and Pd is close to 2.75 A., more than the 2.3 A generally accepted as chemisorption and much less than the 3.35 A needed to characterize physisorption. The graphene bond to Pd (-11 kJ/mol) is significantly smaller than the graphene bond to Ni (-15 kJ/mol), hence easier to break. Nonetheless, it is still large in comparison with the G-Pt adsorption energy (-6 kJ/mol). Notwithstanding the indefinition between physisorption and chemisorption, the Pd surface is an attractive candidate to support graphene-monolayer growth, because it presents remarkable advantages over other metals surfaces. Compare it, for instance, to Pt support. Unlike Pt, palladium allows C growth parallel to the substrate surface, a well structured graphene layer being formed with a distinct Moiré pattern [24]. Meanwhile, in graphene adsorption on Pt the carbon vacancies play an important role and corrugated graphene islands are systematically formed [25]. Compared to Ni, Pd support is also preferable, because the graphene monolayer is more easily extracted from the latter, given that the binding energy for G-Pd (-11 kJ/mol) is smaller than the G-Ni energy (-15 kJ/mol).

We hope that these results encourage the experimental research of graphite deposition on Pd, in analogy with Novoselov's study of deposition on Si [1].

**Acknowledgements**

The author is indebted to the Departamento de Fisica of UNS, IFISUR-CONICET for their financial support and to Mr. and Mrs. Owen for checking the English spelling and grammar.